\documentclass[12pt]{iopart}
\usepackage{graphicx}

\begin{document}

\title[Coherent States of su(1,1)]{Coherent States of su(1,1): Correlations, Fluctuations, and the Pseudoharmonic Oscillator}

\author{John Schliemann}

\address{Institute for Theoretical Physics, University of Regensburg,
D-93040 Regensburg, Germany}
\ead{john.schliemann@physik.uni-regensburg.de}

\begin{abstract}
We extend recent results on expectation values of coherent oscillator
states and SU(2) coherent states to the case of the
discrete representations of su(1,1). Systematic semiclassical expansions
of products of arbitrary operators are derived.
In particular, the leading order of the 
energy uncertainty of an arbitrary Hamiltonian is found to be given purely 
in terms of the time dependence of the classical variables.
The coherent states considered here
include the Perelomov-Gilmore coherent states. As an important application we
discuss the pseudoharmonic oscillator and compare the Perelomov-Gilmore
states with the states introduced by Barut and Girardello. The latter ones
turn out to be closer to the classical limit as their relative energy
variance decays with the inverse square root of energy, while in the former
case a constant is approached.
\end{abstract}

\section{Introduction}
\label{intro}

Coherent states are instrumental in semiclassical descriptions of generic
quantum systems and have proven to be a versatile tool in a plethora
of physical problems. The most prominent types of coherent states
\cite{Klauder85,Perelomov86,Zhang90,Barnett97,Gazeau99,Gazeau09} 
are the coherent
states of the harmonic oscillator, already investigated by Schr\"odinger
\cite{Schrodinger26,Glauber63}, 
and SU(2) coherent states living in the Hilbert space 
of a spin of general length $S$
\cite{Radcliffe71,Arecchi72}.

In both cases the corresponding coherent states fulfill a list of wellknown 
properties which are the basis for their prominent role in semiclassics: 
(i) The coherent
states can be generated by a group transformation from an appropriate 
reference state, (ii) they are  (over-)complete, and  
(iii) are eigenstates of simple operators generic to the system. Moreover, (iv)
they saturate uncertainty relations with respect to an obvious choice of
variables, and (v) they show a coherent time evolution
perfectly mimicking the classical limit under appropriate Hamiltonians.

Recently the present author has argued that one can add to the above list
very general results on the coherent expectation values of products
of arbitrary operators \cite{Schliemann15}. In particular, the leading-order
contribution to the energy uncertainty was, for an arbitrary Hamiltonian, 
found to be purely given by the time dependence of the classical variables,
a both very intuitive and very general result. 
Preliminary findings in this direction appeared also earlier and were obtained
using concretely specified Hamiltonians \cite{Schliemann98,Schliemann99}.
One of the purposes of the present note is to generalize the results
of Ref.~\cite{Schliemann15} to the case of the discrete representations
of su(1,1). As we shall see below, such a generalization is possible for
coherent states of the Perelomov-Gilmore (PG) type
\cite{Perelomov72,Gilmore72,DAriano85}, but not for 
Barut-Girardello (BG) coherent states of su(1,1) \cite{Barut71}. 

A very natural physical system to be studied in connection with the
algebra su(1,1) is the pseudoharmonic oscillator, see e.g.
Refs.~\cite{Moshinsky72,Dodonov74,Agarwal95,Fu96,Dodonov98,Leach08,Mojaveri13,Tavassoly13}.
Most recently, Zipfel and Thiemann \cite{Zipfel15} have revisited this model
under the aspect of {\em complexifier coherent states}
\cite{Thiemann06}, 
a concept originally
inspired by Loop Quantum Gravity \cite{Thiemann07}, see also
Refs.~\cite{Sahlmann01,Thiemann01a,Thiemann01b,Thiemann01c}. As shown in 
Ref.~\cite{Zipfel15}, demanding a complexifier coherent state to have
a stable time evolution (in the sense of the above property (v)) is quite
restrictive, and for one-dimensional systems (described by a single pair
of canonical variables) the only two realizations 
are the usual harmonic oscillator
and the pseudoharmonic oscillator. In the latter case the corresponding
coherent states are of the BG type. For a deeper analysis regarding the
most general Hamiltonian generating stable time evolutions of PG coherent
states, we refer to Ref.~\cite{DAriano85}.
In the present note
we indeed compare PG and BG coherent states
for the pseudoharmonic oscillator in terms of their energy
expectation values and pertaining variances. As a result, the
BG coherent states
turn out to be closer to the classical limit as their relative energy
variance decays with the inverse square root of energy, while for
PG coherent states a constant is approached.

This paper is organized as follows: In section \ref{general} we summarize
important properties of SU(1,1) and its algebra. In particular, we
derive explicit matrix representations of finite transformations 
as applied to su(1,1) operators. Using these findings we construct
in section \ref{costates} a family of coherent states which includes the
PG states. The BG coherent states are also
introduced here. The results about expectation values of products of
arbitrary operators within su(1,1) coherent states are derived and discussed
in section \ref{corrfluc}. Section \ref{pseudohar} is devoted to the
pseudoharmonic oscillator. We close with a summary and an outlook in
section \ref{sumout}.

\section{SU(1,1): General Properties}
\label{general}

The Lie algebra su(1,1) is generated by three operators $K^i$, $i\in\{1,2,3\}$,
fulfilling the commutation relations
\begin{equation}
\left[K^i,K^j\right]=i\epsilon^{ijk}\eta_{kl}K^l
\label{commrel1}
\end{equation}
where summation over repeated indices is understood. The metric 
$\eta_{ij}=\eta^{ij}={\rm diag}(1,1,-1)$ will in the following raise and lower 
indices, and the global sign of the totally antisymmetric tensor
$\epsilon^{ijk}$ is defined by $\epsilon^{123}=+1$. In terms of the usual
complex combinations $K^{\pm}=K^1\pm iK^2$ the above relations can also be
formulated as
\begin{equation}
\left[K^3,K^{\pm}\right]=\pm K^{\pm}\qquad,\qquad
\left[K^+,K^-\right]=-2K^3\,.
\label{commrel2}
\end{equation}
All generators commute with the Casimir invariant
\begin{equation}
C=-K_iK^i=-\frac{1}{2}\left(K^+K^-+K^-K^+\right)+K^3K^3\,.
\label{defC}
\end{equation}
Elements of the pseudounitary group SU(1,1) are obtained by exponentiation,
\begin{equation}
U(\tau,n)=e^{i\tau n_iK^i}
\label{trafo1}
\end{equation}
with a real parameter $\tau$ and and a real unit vector $n^i$ which can 
either be ``spacelike'', $n_in^i=+1$, or ``timelike'', $n_in^i=-1$.
Evaluating the expansion
\begin{equation}
e^XYe^{-X}=\sum_{m=0}^{\infty}\frac{1}{m!}\left[X,Y\right]_m
\label{itcom}
\end{equation}
with $[X,Y]_0=Y$ and $[X,Y]_m=[X,[X,Y]_{m-1}]$,
one finds
\begin{equation}
\tilde K^i:=e^{i\tau n_jK^j}K^ie^{-i\tau n_kK^k}={\left(M(\tau,n)\right)^i}_jK^j
\label{trafo2}
\end{equation}
where the matrix on the r.h.s. is given for spacelike unit vectors
$n$ as
\begin{equation}
{\left(M(\tau,n)\right)^i}_j=n^in_j+\left(-n^in_j+\delta^i_j\right)\cosh\tau
-{{\epsilon^i}_j}^kn_k\sinh\tau
\label{M1}
\end{equation}
while for timelike $n$ we have
\begin{equation}
{\left(M(\tau,n)\right)^i}_j=-n^in_j+\left(n^in_j+\delta^i_j\right)\cos\tau
-{{\epsilon^i}_j}^kn_k\sin\tau\,.
\label{M2}
\end{equation}
In both cases these matrices are elements of the pseudoorthogonal group
O(2,1),
\begin{equation}
{M^i}_k\eta^{kl}{\left(M^T\right)_l}^j=\eta^{ij}
\end{equation}
and the inverses are obtained by inverting either the sign of $\tau$
or $n$,
\begin{equation}
{\left(M^{-1}(\tau,n)\right)^i}_j={\left(M(-\tau,n)\right)^i}_j
={\left(M(\tau,-n)\right)^i}_j\,.
\end{equation}
By construction the transformation (\ref{trafo1}) leave the
commutation relation (\ref{commrel1}) invariant,
\begin{equation}
\left[\tilde K^i,\tilde K^j\right]
=e^{i\tau n_jK^j}\left[K^i,K^j\right]e^{-i\tau n_kK^k}
=i\epsilon^{ijk}\eta_{kl}\tilde K^l=i{\epsilon^{ij}}_k{M^k}_lK^l\,.
\label{commrel3}
\end{equation}
In what follows we will focus on unitary representations of SU(1,1), i.e.
those where all generators $K^i$ are hermitian such that the group
elements (\ref{trafo1}) are unitary. Specifically we focus on the
{\em discrete series} where one
can concentrate here on the ascending series as the descending one
can be treated in a very similar fashion \cite{Barut71,Perelomov86}.
These representations are labeled by a
real parameter $k>0$, and the Hilbert space is of countably 
infinite dimension and spanned by the orthonormalized 
states $|k,m\rangle$, $m\in\{0,1,2,\dots\}$ fulfilling
\begin{eqnarray}
C|k,m\rangle & = & k(k-1)|k,m\rangle\,,
\label{rep1}\\
K^3|k,m\rangle & = & (k+m)|k,m\rangle\,,
\label{rep2}\\
K^+|k,m\rangle & = & \sqrt{(m+1)(2k+m)}|k,m+1\rangle\,,
\label{rep3}\\
K^-|k,m\rangle & = & \sqrt{m(2k-1+m)}|k,m-1\rangle\,.
\label{rep4}\\
\end{eqnarray}
In particular, $K^-|k,0\rangle=0$ and all higher states $|k,m\rangle$, $m>0$
are obtained by applying the raising operator $K^+$.

\section{Coherent States}
\label{costates}

Starting from the lowest-weight state $|k,0\rangle$ one constructs 
via the transformation (\ref{trafo1}) the
family of states
\begin{equation}
|\tau,n\rangle=U(\tau,n)|k,0\rangle
\label{defcssu11-1}
\end{equation}
which fulfill according to Eqs.~(\ref{trafo2}),(\ref{rep2}) 
\begin{equation}
s_iK^i|\tau,n\rangle=k|\tau,n\rangle
\label{defcssu11-2}
\end{equation}
with the timelike unit vector
\begin{equation}
s_i(\tau,n)={\left(M(\tau,n)\right)^3}_i\qquad,\qquad s_is^i=-1\,.
\label{defcssu11-3}
\end{equation}
Eq.~(\ref{defcssu11-2}) strongly resembles a defining property 
of SU(2) coherent states \cite{Radcliffe71,Schliemann98,Schliemann15}
and can therefore be viewed as coherent states of (the discrete series
of representations of) su(1,1). 

From Eqs.~(\ref{defcssu11-2}),(\ref{rep3}),(\ref{rep4}) one easily verifies
that the states (\ref{defcssu11-1}) have the expectation values
\begin{equation}
\langle\tau,n|K^i|\tau,n\rangle=-ks^i
\label{ev1}
\end{equation}
ensuring $\langle\tau,n|s_iK^i|\tau,n\rangle=k$, and for products of
generators one finds
\begin{eqnarray}
\left\langle\tau,n\left|\left(e_iK^i\right)^2\right|\tau,n\right\rangle
& = & \left(\langle\tau,n|e_iK^i|\tau,n\rangle\right)^2\nonumber\\
& & -\frac{1}{2k}\eta_{ij}\langle\tau,n|[e_kK^k,K^i]|\tau,n\rangle
\langle\tau,n|[e_lK^l,K^j]|\tau,n\rangle\nonumber\\
& = & k^2(e_is^i)^2+\frac{k}{2}\epsilon^{ikm}e_ks_m\epsilon_{iln}e^ls^n\nonumber\\
& = & k^2(e_is^i)^2+\frac{k}{2}\left(e_ie^i+(e_is^i)^2\right)
\label{ev2}
\end{eqnarray}
for some arbitrary space- or timelike unit vector $e_i$. A simple way to prove
Eq.~(\ref{ev2}) is to observe that it is fulfilled for the lowest-weight
state $|0,n\rangle=|k,0\rangle$,
\begin{eqnarray}
\left\langle k,0\left|\left(e_iK^i\right)^2\right|k,0\right\rangle
& = & \left(\langle k,0|e_iK^i|k,0\rangle\right)^2\nonumber\\
& & -\frac{1}{2k}\eta_{ij}\langle k,0|[e_kK^k,K^i]|k,0\rangle
\langle k,0|[e_lK^l,K^j]|k,0\rangle\nonumber\\
 & = & \left(ke^3\right)^2
+\frac{k}{2}\left(\left(e^1\right)^2+\left(e^2\right)^2\right)\,,
\end{eqnarray}
and by inserting the transformation (\ref{trafo1}) and its inverse in the
above l.h.s. it follows with the help of Eqs.~(\ref{commrel3}),(\ref{ev1})
\begin{eqnarray}
\left\langle\tau,n\left|\left(e_i\tilde K^i\right)^2\right|\tau,n\right\rangle
& = & 
\left\langle\tau,n\left|\left(e_i{M^i}_jK^j\right)^2\right|\tau,n\right\rangle
\nonumber\\
& = & \left(\langle\tau,n|e_i\tilde K^i|\tau,n\rangle\right)^2\nonumber\\
& & -\frac{1}{2k}\eta_{ij}\langle\tau,n|[e_k\tilde K^k,\tilde K^i]|\tau,n\rangle
\langle\tau,n|[e_l\tilde K^l,\tilde K^j]|\tau,n\rangle\nonumber\\
& = & k^2(e_i{M^i}_js^j)^2+\frac{k}{2}\left(e_ie^i+(e_i{M^i}_js^j)^2\right)\,.
\end{eqnarray}
Eq.~(\ref{ev2}) is now obtained by shifting the arbitrary unit vector
as $e_i\mapsto e_k{(M^{-1})^k}_i$.
As a consequence, the variances squared of such operators read
\begin{eqnarray}
\left(\Delta\left(e_iK^i\right)\right)^2 & = & 
\left\langle\tau,n\left|\left(e_iK^i\right)^2\right|\tau,n\right\rangle
-\left(\langle\tau,n|e_iK^i|\tau,n\rangle\right)^2\nonumber\\
& = & \frac{k}{2}\left(e_ie^i+(e_is^i)^2\right)\,,
\label{ev3}
\end{eqnarray}
and two mutually orthogonal spacelike unit vectors $u$, $v$ being 
perpendicular to $s$, $u_is^i=v_is^i=0$ lead to the minimal uncertainty product
\begin{equation}
\Delta\left(u_iK^i\right)\Delta\left(v_iK^i\right)=\frac{k}{2}
=\frac{1}{2}\langle\tau,n|s_iK^i|\tau,n\rangle\,.
\end{equation}
A particular choice for these unit vectors are
$u_i={\left(M(\tau,n)\right)^1}_i$, $v_i={\left(M(\tau,n)\right)^2}_i$.

The su(1,1) coherent states according to Perelomov and Gilmore (PG)
\cite{Perelomov72,Gilmore72,DAriano85,Perelomov86,Zhang90} can now be identified
within the manifold of states (\ref{defcssu11-1}) by choosing for $n^i$ any
spacelike unit vector with $n^3=0$. A convenient parametrization is given by
$n^i=(\sin\varphi,-\cos\varphi,0)$ leading to the transformation operators
\cite{Perelomov86,Zhang90,Barnett97}
\begin{equation}
U(\tau,\phi)=e^{i\tau(\sin\varphi K^1-\cos\varphi K^2)}
=e^{-\bar zK^+}e^{\eta K^3}e^{zK^-}
\label{defPG1}
\end{equation}
with
\begin{equation}
z=\tanh\frac{\tau}{2}e^{i\phi}\qquad,\qquad\eta=2\ln\cosh\frac{\tau}{2}
\label{defPG2}
\end{equation}
such that the PG coherent states read
\begin{equation}
|\Phi(z)\rangle=\left(1-|z|^2\right)^k\sum_{m=0}^{\infty}
\sqrt{\frac{\Gamma\left(2k+m\right)}{m!\Gamma\left(2k\right)}}
(-\bar z)^m|k,m\rangle\,.
\label{defPG3}
\end{equation}
These states fulfill Eq.~(\ref{defcssu11-2}) with
\begin{eqnarray}
s_i & = & \left(\sinh\tau\cos\varphi,\sinh\tau\sin\varphi,\cosh\tau\right)\\
& = & \left(\frac{2{\rm Re}\,z}{1-|z|^2},\frac{2{\rm Im}\,z}{1-|z|^2},
\frac{1+|z|^2}{1-|z|^2}\right)\,.
\end{eqnarray}
A different type of su(1,1) coherent states has been introduced by
Barut and Girardello (BG) \cite{Barut71}. These states are defined to be
eigenstates of the lowering operator,
\begin{equation}
K^-|\Psi(w)\rangle=w|\Psi(w)\rangle
\label{defBG1}
\end{equation}
with some complex eigenvalue $w$. In the standard basis used so far the BG 
coherent states can be formulated as
\begin{equation}
|\Psi(w)\rangle=N(w,k)\sum_{m=0}^{\infty}
\frac{w^m}{\sqrt{m!\Gamma\left(2k+m\right)}}|k,m\rangle
\label{defBG2}
\end{equation}
where the normalization factor
\begin{equation}
N(w,k)=\left(|w|^{-2k+1})I_{2k-1}(2|w|\right)^{-1/2}\,.
\label{defBG3}
\end{equation}
can be expressed in terms of modified Bessel functions,
\begin{equation}
I_{\nu}(x)=\sum_{m=0}^{\infty}\frac{1}{m!\Gamma(\nu+1+m)}
\left(\frac{x}{2}\right)^{2m+\nu}\,.
\label{modbessel1}
\end{equation}
These states are clearly different from the PG coherent states since
an inspection of the equation
\begin{equation}
t_iK^i|\Psi(w)\rangle=\kappa(w)|\Psi(w)\rangle
\label{nogo}
\end{equation}
shows that the only solutions are given by $t_i\propto (1,-i,0)$, 
$\kappa(w)\propto w$ reproducing Eq.~(\ref{defBG1}). In particular 
there is no solution with a real and timelike $t_i$ as demanded by
Eq.~(\ref{defcssu11-2}).
Finally, as the BG coherent states are eigenstates of $K^-=K^1+iK^2$
it is easy to see that the minimize the uncertainty product
\cite{Zipfel15}
\begin{equation}
\Delta_{\rm BG}(K^1)\Delta_{\rm BG}(K^2)
=\frac{1}{2}\langle\Psi(w)|K^3|\Psi(w)\rangle\,.
\end{equation}

\section{Correlations and Fluctuations}
\label{corrfluc}

Let us now consider two operators $A$, $B$ being functions of the generators
$K^i$. Using the completeness of the basis states $|k,m\rangle$, the 
expectation value of the operator product $AB$ within the states
$|\tau,n\rangle$ can be formulated as
\begin{eqnarray}
\left\langle\tau,n\left|AB\right|\tau,n\right\rangle & = & 
\sum_{m=0}^{\infty}\left\langle k,0\left|U^+AU\right|k,m\right\rangle
\left\langle k,m\left|U^+BU\right|k,0\right\rangle\nonumber\\
& = & \sum_{m=0}^{\infty}\frac{\Gamma(2k)}{m!\Gamma(2k+m)}\Biggl[
\left\langle k,0\left|\left[iK^+,U^+AU\right]_m
\right|k,0\right\rangle\nonumber\\
& & \qquad\qquad\left
\langle k,0\left|\left[iK^-,U^+BU\right]_m\right|k,0\right\rangle
\Biggr]\nonumber\\
& = & \sum_{m=0}^{\infty}\frac{\Gamma(2k)}{m!\Gamma(2k+m)}\Biggl[
\left\langle\tau,n\left|\left[i\tilde K^+,A\right]_m
\right|\tau,n\right\rangle\nonumber\\
& & \qquad\qquad\left
\langle \tau,n\left|\left[i\tilde K^-,B\right]_m\right|\tau,n\right\rangle
\Biggr]
\label{expansion1}
\end{eqnarray}
where $\tilde K^{\pm}$ are given by Eq.~(\ref{trafo2}).
The above last equation extends results of Ref.~\cite{Schliemann15},
obtained there for harmonic oscillator coherent states and SU(2) coherent
states, to the case of the discrete representations of su(1,1).
All the iterated commutators in Eq.~(\ref{expansion1}) are of the same order 
in $k$ whereas the
prefactor of the $m$-th term carries a product $2k(2k+1)\cdots(2k-1+m)$ in the
denominator. Thus, Eq.~(\ref{expansion1}) is essentially an expansion in $1/k$,
and in situations where the classical limit is approached via $k\to\infty$,
it is therefore a systematic semiclassical expansion of the 
coherent-state expectation value of a product of two arbitrary operators. 
These operators are so far neither required to be hermitian nor commuting,
and a different ordering would exchange the operators
$\tilde K^{\pm}$ in Eq.~(\ref{expansion1}) which in general describes
a complex number.

An
example for such a form of the classical limit is given by the pseudoharmonic
oscillator to be discussed in section \ref{pseudohar}.  The zeroth order 
in Eq.~(\ref{expansion1}) is obviously just the classical result.
Note also that the su(1,1) generators 
$\tilde K^x$, $\tilde K^y$ represent the direction perpendicular to the
polarization $s^i$ of the coherent state $|\tau,n\rangle$.
Moreover, for the variance of an hermitian operator $A$ one finds
\begin{equation}
\left(\Delta A\right)^2=\sum_{m=1}^{\infty}\frac{\Gamma(2k)}{m!\Gamma(2k+m)}
\left|\left\langle\tau,n\left|
\left[i\tilde K^-,A\right]_m\right|\tau,n\right\rangle\right|^2\,.
\label{expansion2}
\end{equation}
The expectation values occurring in leading order can be rewritten as
\begin{eqnarray}
\left|\left\langle\tau,n\left|\left[i\tilde K^-,A\right]
\right|\tau,n\right\rangle\right|^2 & = &
\eta_{ij}\left\langle\tau,n\left|\left[i\tilde K^i,A\right]
\right|\tau,n\right\rangle
\left\langle\tau,n\left|\left[i\tilde K^j,A\right]
\right|\tau,n\right\rangle\nonumber\\
& = &
\eta_{ij}\left\langle\tau,n\left|\left[iK^i,A\right]
\right|\tau,n\right\rangle
\left\langle\tau,n\left|\left[iK^j,A\right]
\right|\tau,n\right\rangle\,,
\label{commid}
\end{eqnarray}
where we have observed that $|\tau,n\rangle$ is an eigenstate of $\tilde K^z$,
and that $\tilde K^i$ and $K^i$ are related by an pseudoorthogonal matrix
preserving the metric $\eta_{ij}$.
Thus, we have
\begin{equation}
\left(\Delta A\right)^2=\frac{1}{2k}
\eta_{ij}\left\langle\tau,n\left|\left[iK^i,A\right]
\right|\tau,n\right\rangle
\left\langle\tau,n\left|\left[iK^j,A\right]
\right|\tau,n\right\rangle\nonumber
+{\cal O}\left(\frac{1}{k^2}\right)\,,
\label{expansion3}
\end{equation}
and choosing $A={\cal H}$ to be the Hamiltonian of some system, we
can formulate the leading-order contribution to the energy variance
as
\begin{equation}
\left(\Delta{\cal H}\right)^2=\frac{1}{2k}
\eta_{ij}\left\langle\tau,n\left|\partial_tK^i\right|\tau,n\right\rangle
\left\langle\tau,n\left|\partial_tK^j\right|\tau,n\right\rangle
+{\cal O}\left(\frac{1}{k^2}\right)\,,
\label{expansion4}
\end{equation}
where the commutators have been replaced, according to the Heisenberg
equations of motion, with time derivatives ($\hbar=1$).
Thus, if the system is prepared at some initial time 
in a coherent state $|\tau,n \rangle$ Eq.~(\ref{ev1}) implies
\begin{equation}
\left(\Delta{\cal H}\right)^2=k^2\left(\frac{1}{2k}
\eta_{ij}\left\langle\tau,n\left|\partial_ts^i\right|\tau,n\right\rangle
\left\langle\tau,n\left|\partial_ts^j\right|\tau,n\right\rangle
+{\cal O}\left(\frac{1}{k^2}\right)\right)\,,
\label{expansion5}
\end{equation}
i.e. the leading-order contribution to the energy variance is just due to
the time-dependence of the (semi-)classical coherent parameters.
The results (\ref{expansion1}),(\ref{expansion2}) and
(\ref{expansion3})-(\ref{expansion5}) are in full analogy to the findings
of Ref.~\cite{Schliemann15} for the coherent states of the harmonic 
oscillator and SU(2). Moreover, for a Hamiltonian being linear in the
su(1,1) generators, the energy uncertainty is, according to Eq.(\ref{ev3}),
just given by the leading order in Eq.~(\ref{expansion5}), without any
further correction. This observation is also in full analogy with the
findings of Ref.~\cite{Schliemann15}, and the pseudoharmonic oscillator
to be discussed in section \ref{pseudohar} provides an example for such a
situation.

On the other hand, the above derivation leading to 
Eqs.~(\ref{expansion3})-(\ref{expansion5}) cannot be repeated for
BG coherent states because these objects fail to be
generated via SU(1,1) transformations from the lowest-weight state
$|k,0\rangle$, as seen in Eq.~(\ref{nogo}). Indeed, a unitary
transformation $V(w)$ with
\begin{equation}
|\Psi(w)\rangle=V(w)|k,0\rangle
\end{equation}
is necessarily not an element of the pertaining representation of SU(1,1),
i.e. $V(w)$ is not the form (\ref{trafo1}). As a consequence, there is
still an analog of Eq.~(\ref{expansion1}),
\begin{eqnarray}
\left\langle\Psi(w)\left|AB\right|\Psi(w)\right\rangle & = & 
 \sum_{m=0}^{\infty}\frac{\Gamma(2k)}{m!\Gamma(2k+m)}\Biggl[
\left\langle\Psi(w)\left|\left[i\tilde K^+,A\right]_m
\right|\Psi(w)\right\rangle\nonumber\\
& & \qquad\qquad\left
\langle\Psi(w)\left|\left[i\tilde K^-,B\right]_m\right|\Psi(w)\right\rangle
\Biggr]\,,
\label{expansion6}
\end{eqnarray}
and, in turn, of Eq.~(\ref{expansion2}) 
with $\tilde K^i=VK^iV^+$, but a relation of the form (\ref{commid})
does not hold. 

\section{The Pseudoharmonic Oscillator}
\label{pseudohar}

The pseudoharmonic oscillator
\begin{equation}
H=\frac{p^2}{2\mu}+\frac{\mu\omega^2}{2}q^2+\frac{\lambda}{q^2}
\label{pho}
\end{equation}
describes a particle of mass $\mu$ with coordinate $q$ in a potential
whose harmonic part is characterized by a frequency $\omega$ whereas
the parameter $\lambda$ mimics an angular momentum; for the latter fact this
system is also referred to as the radial oscillator
\cite{Moshinsky72,Zipfel15}. Due to the divergence
of the potential at $q=0$ (giving also rise to the term ``singular oscillator''
\cite{Dodonov74,Perelomov86,Dodonov98})
the dynamics can be restricted to $q\geq 0$.
Moreover, the Hamiltonian (\ref{pho}) describes the relative coordinate of 
the Calogero-Sutherland model in the sector of to just two particles
\cite{Agarwal95,Fu96}.
 
Using the ladder operators of the usual harmonic oscillator,
\begin{equation}
a=\frac{1}{\sqrt{2}}\left(\sqrt{\frac{\mu\omega}{\hbar}}q
+\frac{ip}{\sqrt{\hbar\mu\omega}}\right)\,,
\end{equation}
one constructs a representation of su(1,1) as
\cite{Perelomov86,Agarwal95,Fu96,Dodonov98,Zipfel15}
\begin{equation}
K^-=\frac{1}{2}aa-\frac{\lambda}{2\hbar\omega q^2}\quad,\quad
K^+=\frac{1}{2}a^+a^+-\frac{\lambda}{2\hbar\omega q^2}\quad,\quad
K^3=\frac{H}{2\hbar\omega}
\label{phosu11-1}
\end{equation}
with the Casimir operator
\begin{equation}
C=-\frac{3}{16}+\frac{\mu\lambda}{2\hbar^2}=-\frac{1}{4}+\frac{\alpha^2}{16}
\label{phosu11-2}
\end{equation}
where $\alpha^2=8\mu\lambda/ \hbar^2+1$\,. The eigenstates
of $H=2\hbar\omega K^3$ can be worked out in real-space representation
by standard methods giving (assuming $\alpha>0$)
\cite{Moshinsky72,Perelomov86,Zipfel15}
\begin{equation}
\langle q|k,m\rangle=\sqrt{\sqrt{\frac{\mu\omega}{\hbar}}
\frac{2m!}{\Gamma\left(\frac{\alpha}{2}+1+m\right)}}
e^{-\frac{\mu\omega}{2\hbar}q^2}
\left(\sqrt{\frac{\mu\omega}{\hbar}}q\right)^{\frac{\alpha+1}{2}}
L^{\alpha/2}_m\left(\frac{\mu\omega}{\hbar}q^2\right)
\label{eigenf}
\end{equation}
where
\begin{equation}
L^{\alpha}_m(x)=\left(
\begin{array}{c}
m+\alpha \\
m
\end{array}
\right)
F(-m,\alpha+1,x)
\end{equation}
is a generalized Laguerre polynomial expressed here in terms of
Kummer\rq s function \cite{Abramowitz65}. 
These states fulfill the stationary Schr\"odinger equation
\begin{equation}
H|k,m\rangle=2\hbar\omega\left(m+\frac{\alpha}{4}+\frac{1}{2}\right)|k,m\rangle
\end{equation}
showing that the above su(1,1) representation carries
\begin{equation}
k=\frac{\alpha}{4}+\frac{1}{2}
=\frac{1}{2}+\sqrt{\frac{\mu\lambda}{2\hbar^2}+\frac{1}{16}}\,,
\label{phok}
\end{equation}
consistent with Eq.~(\ref{phosu11-2}). Note that the classical limit
$\hbar\to 0$ implies $k\to\infty$ with
\begin{equation}
\lim_{\hbar\to 0}\hbar k=\sqrt{\mu\lambda/2}\,,
\label{phocl}
\end{equation}
very similar to the classical limit of SU(2) spin systems 
\cite{Schliemann98,Schliemann15}.

The classical dynamics of the variable $q$ is clearly restricted to either
the positive or the negative axis due to the diverging potential barrier
at $q=0$. Accordingly, the wave functions (\ref{eigenf}) yield for
$\lambda=0$ only the odd eigenstates of the usual harmonic oscillator
which vanish at $q=0$ and have energy $\hbar\omega(2m+1+1/2)$
\cite{Dodonov74,Zipfel15}. The even
states are contained in wave functions obtained by changing
$\alpha\mapsto -\alpha$ in Eqs.~(\ref{eigenf})-(\ref{phok}) (but still
assuming $\alpha>0$) leading to \cite{Dodonov74,Leach08}
\begin{equation}
\langle q|k^{\prime},m\rangle=\sqrt{\sqrt{\frac{\mu\omega}{\hbar}}
\frac{2m!}{\Gamma\left(-\frac{\alpha}{2}+1+m\right)}}
e^{-\frac{\mu\omega}{2\hbar}q^2}
\left(\sqrt{\frac{\mu\omega}{\hbar}}q\right)^{\frac{-\alpha+1}{2}}
L^{-\alpha/2}_m\left(\frac{\mu\omega}{\hbar}q^2\right)
\label{eigenfprime}
\end{equation}
with
\begin{equation}
k^{\prime}=-\frac{\alpha}{4}+\frac{1}{2}
=\frac{1}{2}-\sqrt{\frac{\mu\lambda}{2\hbar^2}+\frac{1}{16}}\,.
\label{phokprime}
\end{equation}
These wave functions diverge at $q=0$ for $\lambda>0$ (i.e. $\alpha>1$)
but are still normalizable if $\alpha<2\Leftrightarrow k>0$. Due to the
latter restriction these states do not allow for a classical limit
$\hbar\to 0$.
According to Eqs.~(\ref{phok}),(\ref{phokprime}) the states (\ref{eigenf})
and (\ref{eigenfprime}) form inequivalent representations of su(1,1).
If not stated otherwise we will
in what follows focus on the regular eigenstates (\ref{eigenf}).

On the other hand, integrating the classical energy conservation law
\begin{equation}
E_{\rm cl}=\frac{\mu}{2}\dot q^2_{\rm cl}
+\frac{\mu\omega^2}{2}q^2_{\rm cl}+\frac{\lambda}{q^2_{\rm cl}}
\end{equation}
one finds the general classical solution \cite{Agarwal95,Zipfel15}
\begin{equation}
q_{\rm cl}(t)=\sqrt{\frac{E_{\rm cl}}{\mu\omega^2}+\eta\left(E_{\rm cl}\right)
\cos\left(2\omega t+\varphi\right)}
\end{equation}
with
\begin{equation}
\eta\left(E_{\rm cl}\right)=\sqrt{\left(\frac{E_{\rm cl}}{\mu\omega^2}\right)^2
-\frac{2\lambda}{\mu\omega^2}}
\end{equation}
and $\varphi$ being determined by the initial condition. 
We note that the classical energy is bounded from below by its minimum 
$E_{\rm cl}^{\rm min}=\sqrt{2\mu\lambda\omega^2}$.

Under the quantum Hamiltonian (\ref{pho}) the PG coherent states constructed
from the regular eigenstates (\ref{eigenf}) evolve as 
\begin{equation}
e^{-\frac{i}{\hbar}Ht}|\Phi(z)\rangle=e^{-i2\omega kt}|\Phi(ze^{i2\omega t})\rangle
=:e^{-i2\omega kt}|\Phi(z(t))\rangle
\end{equation}
and remain therefore on the manifold of PG coherent states, i.e. they
are {\em stable} in the sense of Refs.~\cite{DAriano85,Zipfel15}. This property
is completely analogous to the time evolution of the coherent states of the
usual harmonic oscillator and SU(2) coherent states under appropriate
Hamiltonians \cite{Klauder85,Perelomov86,Zhang90,Barnett97,Gazeau99,Gazeau09,Schliemann15,Schliemann98,Schliemann99}. To make further contact with the classical
dynamics we investigate the expectation values of the ``transversal''
su(1,1) components
\begin{equation}
K^1=\frac{\mu\omega}{2\hbar}q^2-\frac{H}{2\hbar\omega}\qquad,\qquad
K^2=\frac{-1}{4\hbar}\left(qp+pq\right)\,.
\end{equation}
According to Eqs.~(\ref{defcssu11-2}),(\ref{M2}) the PG coherent states fulfill
\begin{equation}
s_i(t)K^i|\Phi(z(t))\rangle=k|\Phi(z(t))\rangle
\end{equation}
with
\begin{eqnarray}
s_i(t) & = & \left(\sinh\tau\cos(2\omega t+\varphi),
\sinh\tau\sin(2\omega t+\varphi),\cosh\tau\right)\\
& = & \left(\frac{2{\rm Re}\,z(t)}{1-|z|^2},\frac{2{\rm Im}\,z(t)}{1-|z|^2},
\frac{1+|z|^2}{1-|z|^2}\right)
\end{eqnarray}
such that
\begin{eqnarray}
\langle\Phi(z(t))|K^1|\Phi(z(t))\rangle
& = & \frac{-2k{\rm Re}\,z(t)}{1-|z|^2}\,,
\label{PGK1-1}\\
\langle\Phi(z(t))|K^2|\Phi(z(t))\rangle
& = & \frac{-2k{\rm Im}\,z(t)}{1-|z|^2}\,.
\label{PGK2-1}
\end{eqnarray}
For these time-dependent expectation values to be identical to the
corresponding classical quantities we must have 
\begin{equation}
\langle\Phi(z(t))|K^2|\Phi(z(t))\rangle
=\frac{-m}{2\hbar}\dot q_{\rm cl}(t)q_{\rm cl}(t)
=\frac{m\omega}{2\hbar}\eta\left(E_{\rm cl}\right)
\sin\left(2\omega t+\varphi\right)
\label{PGK2-2}
\end{equation}
leading to 
\begin{equation}
\frac{-2kz(t)}{1-|z|^2}
=\frac{m\omega}{2\hbar}\eta\left(E_{\rm cl}\right)e^{i(2\omega t+\varphi)}
\end{equation}
and 
\begin{equation}
\langle\Phi(z(t))|K^1|\Phi(z(t))\rangle
=\frac{m\omega}{2\hbar}\eta\left(E_{\rm cl}\right)
\cos\left(2\omega t+\varphi\right)
=\frac{m\omega}{2\hbar}q^2_{\rm cl}(t)
-\frac{E_{\rm cl}}{2\hbar\omega}\,.
\label{PGK1-2}
\end{equation}
An analogous observation can be made for BG coherent state where the
time evolution is also stable,
\begin{equation}
e^{-\frac{i}{\hbar}Ht}|\Psi(w)\rangle=e^{-i2\omega kt}|\Psi(we^{-i2\omega t})\rangle
=:e^{-i2\omega kt}|\Psi(w(t))\rangle\,,
\end{equation}
leading to
\begin{eqnarray}
\langle\Psi(w(t))|K^1|\Psi(w(t))\rangle & = & {\rm Re}\,w(t)\,,\\
\langle\Psi(w(t))|K^2|\Psi(w(t))\rangle & = & -{\rm Im}\,w(t)\,.
\end{eqnarray}
Putting now
\begin{equation}
w(t)=\frac{m\omega}{2\hbar}\eta\left(E_{\rm cl}\right)e^{-i(2\omega t+\varphi)}
\end{equation}
we have as before
\begin{eqnarray}
\langle\Psi(w(t))|K^1|\Psi(w(t))\rangle & = & 
\frac{m\omega}{2\hbar}q^2_{\rm cl}(t)
-\frac{E_{\rm cl}}{2\hbar\omega}\,,
\label{BGK1}\\
\langle\Psi(w(t))|K^2|\Psi(w(t))\rangle & = & 
\frac{-m}{2\hbar}\dot q_{\rm cl}(t)q_{\rm cl}(t)\,.
\label{BGK2}
\end{eqnarray}
Both the PG and the BG coherent states of su(1,1) perfectly mimic the classical
dynamics of the pseudoharmonic oscillator. Specifically the moduli of
the complex parameters are to be chosen as
\begin{equation}
\frac{2k|z|}{1-|z|^2}=|w|=\frac{m\omega}{2\hbar}\eta\left(E_{\rm cl}\right)
\end{equation}
such that
\begin{equation}
\frac{-2kz}{1-|z|^2}
=\bar w\quad\Leftrightarrow\quad
z=\frac{1}{w}\left(k-\sqrt{k^2+|w|^2}\right)\,.
\end{equation}
The above observations are of course in close analogy to wellknown
properties of the coherent states of the usual harmonic oscillator and
of SU(2) coherent states \cite{Klauder85,Perelomov86,Zhang90,Barnett97,Gazeau99,Gazeau09,Schliemann15,Schliemann98,Schliemann99}. 
The relationship between PG and BG coherent state to the classical
dynamics was already investigated in Ref.~\cite{Agarwal95} concentrating on
the time-dependence of the modulus of the coherent-state wave functions.
In particular, the imaginary parts of the coherent parameters
$z$ and $\omega$ corresponding to the expectation values of $K^2$ were
not considered. 

On the other hand, we note
that a stable time evolution mimicking the classical limit is as such
not a particularly distinctive property \cite{Zhang90,Gazeau99}. 
As an example consider states of the form
\begin{equation}
|\chi(z)\rangle=M(z)\sum_{m=0}^{\infty}c_mz^m|k,m\rangle
\label{defchi}
\end{equation}
where the complex numbers $c_m$ are chosen such that the series
\begin{equation}
(M(z))^{-1/2}=\sum_{m=0}^{\infty}|c_m|^2|z|^{2m}
\end{equation}
has a finite radius of convergence, but are otherwise arbitrary. Such
states are obviously stable under the Hamiltonian time evolution.
Moreover, let us further assume the expectation value
\begin{equation}
\langle\chi(z)|K^+|\chi(z)\rangle=\frac{\bar z}{|z|}f(|z|)
\end{equation}
with
\begin{equation}
f(|z|)=\sum_{m=1}^{\infty}\bar c_mc_{m-1}\sqrt{m(2k-1+m)}|z|^{2m-1}
\end{equation}
to be also finite. Choosing then $|z|$ according to
\begin{equation}
f(|z|)=\frac{m\omega}{2\hbar}\eta\left(E_{\rm cl}\right)
\end{equation}
leads to expectation values of $K^1$, $K^2$ which have the identical
classical time evolution as in 
Eqs.~(\ref{PGK2-2}),(\ref{PGK1-2}) and (\ref{BGK1}),(\ref{BGK2}).
However, for general 
coefficients $c_m$ the resulting state can certainly not be expected
to have other properties desired for semiclassical
approximations such as minimum uncertainty products as realized by
PG and BG coherent states. Another 
important feature are of course the expectation values of $K^3$ which
we now investigate.

For the expectation value $H=2\hbar\omega K^3$ we have within a PG
coherent state from Eq.~(\ref{ev1})
\begin{equation}
\langle\Phi(z)|H|\Phi(z)\rangle
=2\hbar\omega k\cosh\tau
=\sqrt{E_{\rm cl}^2+(2\hbar\omega k)^2-2\mu\lambda\omega^2}
\end{equation}
which approaches $E_{\rm cl}$ in the semiclassical regime of large energies
$E_{\rm cl}\gg\hbar\omega$, 
$E_{\rm cl}\gg E_{\rm cl}^{\rm min}=\sqrt{2\mu\lambda\omega^2}$. 
The energy variance squared can be calculated
via Eq.~(\ref{ev3}) as
\begin{equation}
\left(\Delta_{\rm PG}H\right)^2=(2\hbar\omega)^2\frac{k}{2}\sinh^2\tau
=\frac{1}{2k}\left(E_{\rm cl}^2-2\mu\lambda\omega^2\right)
\label{PGDeltaH1}
\end{equation}
such that the relative variance approaches a constant at
large energies,
\begin{equation}
\frac{\Delta_{\rm PG}H}{\langle\Phi(z)|H|\Phi(z)\rangle}
=\frac{1}{\sqrt{2k}}+{\cal O}\left(\frac{\hbar\omega}{E_{\rm cl}},
\frac{\sqrt{\mu\lambda\omega^2}}{E_{\rm cl}}\right)\,,
\label{PGDeltaH2}
\end{equation}
which is certainly not the expected behavior for a state incorporating
the semiclassics.

For the BG coherent states we can use the modified Bessel functions
(\ref{modbessel1}) to obtain
\begin{equation}
\langle\Psi(w)|H|\Psi(w)\rangle=2\hbar\omega
\left(k+\frac{|w|I_{2k}(2|w|)}{I_{2k-1}(2|w|)}\right)
\label{BGevH1}
\end{equation}
and
\begin{eqnarray}
\left(\Delta_{\rm BG}H\right)^2 & = & (2\hbar\omega)^2
\Biggl(\frac{|w|^2I_{2k+1}(2|w|)+|w|I_{2k}(2|w|)}{I_{2k-1}(2|w|)}\nonumber\\
& & \qquad\qquad\qquad
-\left(\frac{|w|I_{2k}(2|w|)}{I_{2k-1}(2|w|)}\right)^2\Biggr)\,.
\label{BGDeltaH1}
\end{eqnarray}
Note that the above expression, differently from Eq.~(\ref{PGDeltaH1}),
contains higher orders in $1/k$.

Now employing the asymptotic expansion \cite{Abramowitz65}
\begin{equation}
I_{\nu}(x)=\frac{e^x}{\sqrt{2\pi x}}\left(1+\frac{4\nu^2-1}{8x}
+{\cal O}\left(\frac{1}{x}\right)^2\right)
\label{modbessel2}
\end{equation}
one finds for $E_{\rm cl}\gg\hbar\omega$, $E_{\rm cl}\gg E_{\rm cl}^{\rm min}$ 
\begin{equation}
\langle\Psi(w)|H|\Psi(w)\rangle=E_{\rm cl}+\frac{\hbar\omega}{2}
+{\cal O}\left(\frac{\hbar\omega}{E_{\rm cl}},
\frac{\sqrt{\mu\lambda\omega^2}}{E_{\rm cl}}\right)
\label{BGevH2}
\end{equation}
and
\begin{equation}
\frac{\Delta_{\rm BG}H}{\langle\Psi(w)|H|\Psi(w)\rangle}
=\sqrt{\frac{\hbar\omega}{E_{\rm cl}}}
+{\cal O}\left(\frac{\hbar\omega}{E_{\rm cl}},
\frac{\sqrt{\mu\lambda\omega^2}}{E_{\rm cl}}\right)\,.
\label{BGDeltaH2}
\end{equation}
Thus, the energy expectation value (\ref{BGevH2}) contains
a ``zero-point energy'' $\hbar\omega/2$ very familiar from the standard
harmonic oscillator, while the relative energy variance (\ref{BGDeltaH2})
vanishes in the semiclassical regime with the inverse square root
of energy. The latter property is in contrast to the behavior
(\ref{PGDeltaH2}) of the PG coherent state and an expected feature in the
semiclassical limit.
\begin{figure}
 \includegraphics[width=\columnwidth]{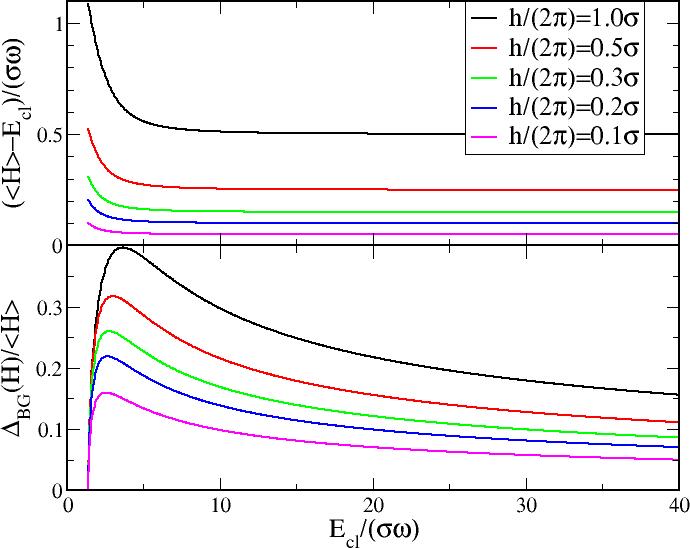}
\caption{The difference 
$\langle\Psi(w)|H|\Psi(w)\rangle-E_{\rm cl}=\langle H\rangle-E_{\rm cl}$
(upper panel, Eq.~(\ref{BGevH1})) and the relative energy uncertainty 
(lower panel, Eqs.~(\ref{BGDeltaH1}),(\ref{BGevH1})) 
for a Barut-Girardello coherent state
as a function of $E_{\rm cl}$ for different values of $\hbar$. All energies are
in units of $\sigma\omega:=\sqrt{\mu\lambda\omega^2}$ while $\hbar$ is
expressed in units of $\sigma=\sqrt{\mu\lambda}$.\\
The data in the upper panel approaches $\hbar\omega/2$ at large energies,
while the relative uncertainty in the lower panel vanishes with the
inverse square root.}
\label{fig1}
\end{figure}
To illustrate the above findings we have plotted the expressions
(\ref{BGevH1}),(\ref{BGDeltaH1}) in Fig.~\ref{fig1} as a function
of $E_{\rm cl}$ for different values of $\hbar$.

The above analysis focused on coherent states constructed from
the regular eigenstates (\ref{eigenf}) of the pseudoharmonic oscillator.
Similarly one could employ the divergent states (\ref{eigenfprime}) which,
however, do not possess a classical limit. More interestingly, as a closer
inspection easily shows, coherent states constructed from either type
of eigenstates, or linear combinations of them with fixed coherent parameters,
do not reproduce for $\lambda=0$ the well-known coherent states of the
usual harmonic oscillator 
\cite{Klauder85,Perelomov86,Zhang90,Barnett97,Gazeau99,Gazeau09}.
The latter statement holds both for PG and BG coherent states.

\section{Summary and Outlook}
\label{sumout}

We have extended recent results \cite{Schliemann15}
on expectation values of operator products within coherent oscillator
states and SU(2) coherent states to the case of the
discrete representations of su(1,1). The results provide a systematic
expansion of correlations and fluctuations around the classical limit.
In particular, the leading order of the energy uncertainty of an arbitrary
Hamiltonian is found, in full analogy to Ref.~\cite{Schliemann15},
to be given purely in terms of the time dependence
of the classical variables. The latter finding holds for a family of 
coherent states including the PG states, but their derivation cannot 
be extended  states to the BG type. Our results regarding PG coherent states
are based on explicit matrix representations of SU(1,1) transformations
derived in section \ref{general}.

As a typical application we have
discussed the pseudoharmonic oscillator and established that the
time evolution of the both the PG and BG coherent states perfectly
mimic, for appropriate choices of the coherent parameters, the
classical dynamics. 
However, departures between these types of coherent states are revealed when
comparing expectation values: While the energy expectation values are
close to each other, the variances show a qualitative difference:
For BG states the relative variance vanishes with the inverse square root
of energy whereas in the PG case a constant is approached.
Thus, in contrast to the PG states, the BG coherent states show a behavior 
perfectly expected in the semiclassical regime. Moreover, the energy
expectation values of BG coherent states contain a zero-point energy
strongly reminiscent of the standard harmonic oscillator.

Possible direction of further work include the extension of the results
obtained in section \ref{corrfluc} and Ref.~\cite{Schliemann15} to
other (compact or noncompact) groups \cite{Perelomov86,Zhang90},
and the study of generalizations of the pseudoharmonic oscillator,
especially to Hamiltonians with explicit time dependence
\cite{Dodonov74,Dodonov98}.

\section*{Acknowledgements}

I thank Antonia Zipfel for useful correspondence.

{}

\end{document}